\documentstyle[12pt,preprint,eqsecnum,aps,prd,epsf]{revtex}

\begin{document}

\draft
\preprint{\begin{tabular}{l}
MADPH-96-947 \\
FERMILAB-PUB-96/144-T
\end{tabular}}

\title{Decay constants of $P$ and $D$-wave heavy-light mesons}

\author{Sini\v{s}a Veseli}
\address{Department of Physics, University of Wisconsin, Madison,
	WI 53706}

\author{Isard Dunietz}
\address{Fermi National Accelerator Laboratory, P.O. Box 500, Batavia,
IL 60510}

\date{\today}

\maketitle

\begin{abstract}
We investigate decay constants
of $P$ and $D$-wave heavy-light mesons within
the mock-meson approach. Numerical estimates are
obtained using  the relativistic quark model. We also
comment on recent calculations of
heavy-light pseudo-scalar
and vector decay constants.
\end{abstract}


\newpage
\section{Introduction}

Reliable estimates of heavy-light meson decay constants
are important, since they appear in many processes from which
fundamental quantities can be  extracted \cite{buchalla}.
Theoretical investigations have focused on estimating
decay constants for the weakly decaying pseudo-scalar meson and its
HQET (heavy quark effective theory) related vector meson.
Whereas the decay constant of the weakly decaying pseudo-scalar
meson is of paramount importance for determining
fundamental quantities,
the decay constant of the $S$-wave vector meson plays a role in
exclusive $b\rightarrow ul\overline{\nu}$ transitions \cite{burdman}
and in radiative leptonic decays of heavy-light mesons
\cite{burdman2}.

While those decay constants have been and continue to be studied
intensively, the decay constants of the more highly
excited heavy-light states have been normally ignored.
This note attempts to rectify this situation, by predicting
decay constants for many higher-excited resonances. That could be
important phenomenologically on several accounts.

First,  CLEO recently
observed a significant wrong charm contribution in $B$ decays
\cite{kwon},
\begin{equation}
{\cal B}(\overline{B}\rightarrow \overline{D}X) \approx 10\%\ ,
\end{equation}
governed essentially by the $b\rightarrow c\overline{c}s'$
quark transitions.\footnote{The prime indicates that the
corresponding
Cabibbo suppressed mode is included.}
The $\overline{B}\rightarrow \overline{D}X$ transitions were
overlooked
in all previous experimental analyses.
Under the factorization assumption \cite{fakirov},
wherein the virtual $W\rightarrow \overline{c}s$ hadronizes
independently to the rest of the system, a quantitative
modelling of the $\overline{B}\rightarrow \overline{D}X$ transitions
can be undertaken once  theory provides the
decay constants for $D_{s}^{**}$.

Second, reliable estimates for decay constants of $D^{**}$ allow one
to test whether color-allowed and color-suppressed
decay amplitudes interfere constructively for the
$B^{-}\rightarrow D^{**0}\{\pi^{-},\rho^{-},a_{1}^{-},\ldots\}$
modes,
as has been seen for the
$B^{-}\rightarrow D^{(*)0}\{\pi^{-},\rho^{-},a_{1}^{-},\ldots\}$
transitions \cite{browder}. Third, such estimates
enable us to better predict subtle CP violating phenomena.

Decay constants are defined through matrix elements of vector and
pseudo-vector currents between meson states and the vacuum.
Therefore, in order to calculate them, one has
to find a way to evaluate hadronic matrix elements.
The mock-meson method \cite{yaou,ruiz,isgu,hayne}
has been frequently used in the literature for that purpose
\cite{yaou,ruiz,isgu,hayne,godf,godf2,grin,isgu2,capstick,scora,hwang}.
In this paper we follow the same approach,
and use the mock-meson method in order to obtain expressions
for the decay constants of heavy-light mesons, in terms
of integrals over momentum-space bound-state wave functions.
For numerical estimates we decided to use
the simplest relativistic generalization of the
Schr$\ddot{\rm o}$dinger equation
\cite{godf,durand,lichtenberg,jacobs},
sometimes called the spinless Salpeter equation.

The rest of the paper is organized as follows. We begin with
a brief description of the mock-meson method in Section \ref{mock}.
Our approach is based on the $j$-$j$ coupling scheme, since
it is more appropriate for heavy-light mesons than the
usual $L$-$S$ scheme. Expressions for the
decay constants of heavy-light meson states
 are given in Section \ref{dcon}. The relativistic
quark model and our  numerical estimates are described
in Section \ref{nres}. There we also comment on recent
calculations of pseudo-scalar and vector decay constants
\cite{hwang}.
Our conclusions are summarized in Section \ref{conc}.

\section{The mock-meson method}
\label{mock}

As already mentioned,
the mock-meson approach \cite{yaou,ruiz,isgu,hayne} has been widely
used for calculations of hadronic matrix elements
\cite{yaou,ruiz,isgu,hayne,godf,godf2,grin,isgu2,capstick,scora,hwang}.
The  basic idea of the method is simple.
The mock meson is defined as a collection of free quarks weighted
with a bound-state wave function.
The mock-meson matrix elements $\widetilde{\cal M}$
can then be calculated  using full Dirac spinors.
On the other hand, the physical matrix elements ${\cal M}$
can always be expressed in terms of Lorentz covariants
with coefficients $A_{i}$, which are Lorentz scalars. In many simple  cases,
${\cal M}$ and $\widetilde{\cal M}$ will be of the same form. The
mock-meson
prescription then says that in those cases one should simply take
$A_{i}=\widetilde{A}_{i}$.
Indeed, this correspondence is exact in the zero-binding limit and
in the meson rest frame. Away from this limit the mock amplitudes
are in general not invariant by terms of order $p_{i}^{2}/m_{i}^{2}$.

In this paper we are primarily concerned with the decay constants
of heavy-light  $q\overline{Q}$ mesons.
In the  $m_{\overline{Q}}\rightarrow \infty$ limit, heavy quark
symmetry tells us that
the angular momentum  of the light degrees of freedom (LDF)
in the heavy-light meson decouples from the spin of the heavy quark,
and both
are separately conserved by the strong interaction \cite{isgu3}.
Therefore, total angular momentum $j$ of the LDF
  is a good quantum number. For
each $j$ there are two
degenerate heavy meson states ($J=j\pm \frac{1}{2}$), which
can be labeled  as $J^{P}_{j}$, where $P=(-1)^{L+1}$.
This implies that in the
case of heavy-light mesons the $j$-$j$ coupling
is more appropriate than the $L$-$S$ coupling scheme.
For this reason, we first define the LDF states
$|j\lambda_{j};L\frac{1}{2}\rangle$ as Clebsch-Gordan (CG)
combinations
of the eigenstates
of orbital angular momentum $|LM_{L}\rangle$,
and those of the spin of the light quark $|\frac{1}{2}s\rangle$,
with CG coefficients denoted as
$C^{j\lambda_{j}}_{LM_{L};\frac{1}{2}s}$. Combining
the LDF states with those of the heavy antiquark
$|\frac{1}{2}\overline{s}\rangle$
(with CG coefficients
$C^{JM_{J}}_{j\lambda_{j};\frac{1}{2}\overline{s}}$),
we get the $q\overline{Q}$ mock meson
state
in its rest frame,
\begin{eqnarray}
|J^{P}_{j}M_{J};n\rangle
 &=& \sqrt{2 \widetilde{M}} \frac{1}{\sqrt{3}}
\sum_{c} \sum_{\lambda_{j},M_{L},s,\overline{s}}
C^{JM_{J}}_{j\lambda_{j};\frac{1}{2}\overline{s}}
C^{j\lambda_{j}}_{LM_{L};\frac{1}{2}s}
\ \times  \nonumber \\
&\times &\int \frac{d^{3}{\bf p}}{(2\pi)^{3}}
 \sqrt{\frac{m_{q}m_{\overline{Q}}}{E_{q}E_{\overline{Q}}}}\phi_{n
LM_{L}}({\bf p})
|q_{c}({\bf p}, s)\rangle |\overline{Q}_{\overline{c}}(-{\bf p},
\overline{s})\rangle\ .
\label{mmes}
\end{eqnarray}
In the above expression $E_{i} = \sqrt{m_{i}^{2}+{\bf p}^{2}}$,
$\widetilde{M}$
is the mock-meson mass, and the color wave function
(subscript c denotes color) is written
explicitly.
Also, $\phi_{n LM_{L}}({\bf p})$ is a normalized
momentum wave function, where $n$ denotes all other quantum numbers
of a state not connected to angular momentum
(e.g.,  radial quantum number). The factor
$\frac{1}{(2\pi)^{3}}\sqrt{\frac{m_{q}m_{\overline{Q}}}
{E_{q}E_{\overline{Q}}}}$
appears due to our normalization convention
for creation and annihilation operators \cite{itzykson},
$\{b_{\alpha}(k), b^{\dagger}_{\alpha'}(k')\} =
(2\pi)^{3}\ \frac{k_{0}}{m}\
\delta^{3}({\bf k} - {\bf k}')\delta_{\alpha\alpha'}$, etc.
The mock-meson states as given in (\ref{mmes})
are normalized to $2\widetilde{M}$.

As already observed in \cite{capstick}, the mock-meson approach
suffers from a number of ambiguities, such as the choice for
quark masses, or
the definition of the mock-meson mass
$\widetilde{M}$. In the spirit of the method, the mock-meson mass
should be defined as $\widetilde{M}=\langle E_{q}\rangle + \langle
E_{\overline{Q}}\rangle$.
However, as pointed out in
\cite{capstick}, the mock-meson mass has been introduced to give the
correct
relativistic normalization of the meson's wave function, and hence
the
use of  the physical meson mass $M$ instead of
$\langle E_{q}\rangle + \langle E_{\overline{Q}}\rangle$ may
be more appropriate. We adopt the same approach, and write
$\widetilde{M}= M$. We also note that
the heavier the mesons are, the less important it is how
the mock-meson mass is defined, since the relativistic effects
and binding energies become less significant. As far as
quark masses are concerned, the self-consistency of the model requires
the use of constituent quark masses.
In our error estimates we have included variations of constituent
light quark masses over a range of about $200\ MeV$, and also of
heavy quark masses over a range of about $400\ MeV$, so that
we believe that uncertainties introduced by a particular
choice of  quark masses are being properly taken into account.

\section{Decay constants}
\label{dcon}

Decay constants of heavy-light mesons are defined through
matrix elements of vector $V^{\mu}$ and pseudo-vector
$A^{\mu}$ currents between a meson state and the vacuum.
Following standard definitions in the literature
\cite{godf}, for pseudo-scalar (P), vector (V), scalar (S), and
pseudo-vector (A) mesons, we write
\begin{eqnarray}
\langle 0 |A^{\mu}(0) | 0_{1/2}^{-}(k)\rangle &=&
\frac{1}{(2\pi)^{3/2}}f_{P} k^{\mu}\ ,
\label{fps}\\
\langle 0 |V^{\mu}(0) | 1_{1/2}^{-}(\epsilon,k)\rangle &=&
\frac{1}{(2\pi)^{3/2}}f_{V_{1/2}} M_{V_{1/2}}\epsilon^{\mu}\ ,
\label{fv12}\\
\langle 0 |V^{\mu}(0) | 0_{1/2}^{+}(k)\rangle &=&
\frac{1}{(2\pi)^{3/2}}f_{S} k^{\mu}\ ,
\label{fs}\\
\langle 0 |A^{\mu}(0) | 1_{1/2}^{+}(\epsilon,k)\rangle &=&
\frac{1}{(2\pi)^{3/2}}f_{A_{1/2}} M_{A_{1/2}}\epsilon^{\mu}\ ,
\label{fpv12}\\
\langle 0 |A^{\mu}(0) | 1_{3/2}^{+}(\epsilon,k)\rangle &=&
\frac{1}{(2\pi)^{3/2}}f_{A_{3/2}} M_{A_{3/2}}\epsilon^{\mu}\ ,
\label{fpv32}\\
\langle 0 |V^{\mu}(0) | 1_{3/2}^{-}(\epsilon,k)\rangle &=&
\frac{1}{(2\pi)^{3/2}}f_{V_{3/2}} M_{V_{3/2}}\epsilon^{\mu}\ .
\label{fv32}
\end{eqnarray}
Note that in the heavy quark limit
$0_{1/2}^{-}$ and  $1_{1/2}^{-}$ states are degenerate
($S$-waves), and so are $0_{1/2}^{+}$ and $1_{1/2}^{+}$
($P$-wave states). The spin 2  members of $P$-wave
($1_{3/2}^{+},2_{3/2}^{+}$) and $D$-wave
($1_{3/2}^{-},2_{3/2}^{-}$) doublets do not
couple leptonically due to conservation of angular momentum.

In order to obtain expressions for decay constants in terms
of integrals over momentum-space meson wave functions,
we  evaluate the matrix elements (\ref{fps})-(\ref{fv32})
in the meson rest frame using (\ref{mmes}).
Of course, any choice of
polarization for spin 1 mesons should yield the same result.
As mentioned earlier, current matrix elements between states
defined in (\ref{mmes}) and the vacuum can be evaluated exactly with
full
Dirac spinors.
Because of  spherical symmetry, the momentum-space
wave function  can be  written in the form
\begin{equation}
\phi_{n LM_{L}}({\bf p}) = R_{nL}(p) Y_{LM_{L}}(\hat{\bf p})\ .
\label{wf}
\end{equation}
In the above $Y_{LM_{L}}$ are the usual spherical harmonics, and $R_{nL}(p)$
is the radial part of the wave function, where $p$ denotes
$|{\bf p}|$ henceforth.
Using (\ref{wf}), and keeping track of the relevant CG coefficients, we
find that all
heavy-light meson decay constants in the mock-meson approach
 can be written in the form
\begin{equation}
f_{i} = \frac{2\sqrt{3}}{\sqrt{M}} \sqrt{4\pi}
\int_{0}^{\infty} \frac{p^{2}dp}{(2\pi)^{3/2}}
\sqrt{ \frac{(m_{q}+E_{q})(m_{\overline{Q}}+E_{\overline{Q}})}{4
E_{q} E_{\overline{Q}} } }
F_{i}(p)\ ,
\label{fi}
\end{equation}
where
\begin{eqnarray}
F_{P}(p) &=& \left[ 1 -
\frac{p^{2}}{(m_{q}+E_{q})(m_{\overline{Q}}+E_{\overline{Q}})}
\right] R_{n0}(p)\ ,
\label{fi1}\\
F_{V_{1/2}}(p) &=& \left[ 1+ \frac{1}{3}
\frac{p^{2}}{(m_{q}+E_{q})(m_{\overline{Q}}+E_{\overline{Q}})}
\right] R_{n0}(p)\ ,
\label{fi2}\\
F_{S}(p) &=& \left[ \frac{1}{(m_{q}+E_{q})} -
\frac{1}{(m_{\overline{Q}}+E_{\overline{Q}})}
\right]p R_{n1}(p)\ ,
\label{fi3}\\
F_{A_{1/2}}(p) &=& \left[ \frac{1}{(m_{q}+E_{q})} +\frac{1}{3}
\frac{1}{(m_{\overline{Q}}+E_{\overline{Q}})}
\right]p R_{n1}(p)\ ,
\label{fi4}\\
F_{A_{3/2}}(p) &=& \left[\frac{2\sqrt{2}}{3}
\frac{1}{(m_{\overline{Q}}+E_{\overline{Q}})}
\right]p R_{n1}(p)\ ,
\label{fi5}\\
F_{V_{3/2}}(p) &=& \left[\frac{2\sqrt{2}}{3}
\frac{1}{(m_{q}+E_{q})}
\frac{1}{(m_{\overline{Q}}+E_{\overline{Q}})}
\right]p^{2} R_{n2}(p)\ .
\label{fi6}
\end{eqnarray}
Expressions (\ref{fi1}) and (\ref{fi2}) were found
in \cite{godf2} and \cite{hwang}, respectively.

It is interesting to observe that in the limit
$m_{\overline{Q}}\rightarrow \infty$ (\ref{fi})-(\ref{fi6}) become
\begin{equation}
f_{i}^{HL} = \frac{2\sqrt{3}}{\sqrt{M}} \sqrt{4\pi}
\int_{0}^{\infty} \frac{p^{2}dp}{(2\pi)^{3/2}}
\sqrt{ \frac{(m_{q}+E_{q})}{2 E_{q} } }
F_{i}^{HL}(p)\ ,
\label{fihl}
\end{equation}
with
\begin{eqnarray}
F_{P}^{HL}(p) &=& F_{V_{1/2}}^{HL}(p) =  R_{n0}(p)\ ,
\label{fi1hl}\\
F_{S}^{HL}(p) &=& F_{A_{1/2}}(p) =
 \frac{1}{(m_{q}+E_{q})} p R_{n1}(p)\ ,
\label{fi2hl}\\
F_{A_{3/2}}^{HL}(p) &=& 0\ ,
\label{fi3hl}\\
F_{V_{3/2}}^{HL}(p) &=& 0\ .
\label{fi4hl}
\end{eqnarray}
Equality of $f_{P}$ and $f_{V_{1/2}}$, and also that of
$f_{S}$ and $f_{A_{1/2}}$, as well as vanishing of
$f_{A_{3/2}}$ and $f_{V_{3/2}}$, are in the heavy quark limit
expected from the heavy quark symmetry.

\section{Relativistic quark model }
\label{nres}

In order to obtain numerical estimates for the
decay constants of heavy-light mesons, we
consider the simplest and widely used generalization
of the non-relativistic Schr$\ddot{\rm o}$\-dinger equation
\cite{godf,durand,lichtenberg,jacobs} with Hamiltonian
given by
\begin{equation}
H = \sqrt{m_{q}^{2} + p^{2}} + \sqrt{m_{\overline{Q}}^{2} + p^{2}}
+ V(r)\ ,
\label{srqm}
\end{equation}
where for $V(r)$ we take the QCD-motivated Coulomb-plus-linear
potential \cite{godf}
\begin{equation}
V(r) = -\frac{4}{3}\frac{\alpha_{s}}{r} + b r  + c\ .
\label{pot}
\end{equation}
For the sake of simplicity,\footnote{The
running coupling constant was used in \cite{godf}.}
we take $\alpha_{s}$ to be a fixed
effective short range coupling constant.
The effective string tension of the model can be determined
from the requirement  that
the linear Regge structure of the model
in the light-light limit agrees with the
observed slope of the $\rho$ trajectory \cite{veseli}.
Fixing $m_{u,d}$, other parameters
can be chosen so that the model reproduces
the observed spin-averaged spectrum of the known
heavy-light states. One such set of parameters includes
constituent quark masses $m_{u,d}=0.300\ GeV$, $m_{s}=0.483\ GeV$,
$m_{c}= 1.671\ GeV$, and $m_{b}=5.121\ GeV$,
and also $\alpha_{s}= 0.498$, $b=0.142\ GeV^{2}$,
and $c=-0.350\ GeV$.\footnote{This particular parameter set
corresponds to the spin-averaged mass of the
unknown $D_{0}$  and $D'_{1}$ mesons
($0^{+}_{1/2}$ and $1^{+}_{1/2}$ states)
of about  $2400\ MeV$.}
As can be seen in Table \ref{spectrum}, these parameters
yield an excellent description of the observed spin-averaged
heavy-light spectrum.

We now  turn to the discussion of  pseudo-scalar and vector decay constants.
Recently, Ref. \cite{hwang} used (\ref{srqm}) with
 six different potentials, and with
current quark masses from \cite{dominguez},
minimized  the Hamiltonian with respect
to the variational parameter $\beta$ of a single harmonic
oscillator (HO) wave function,
\begin{equation}
R_{1S} (p) = \frac{2}{\pi^{1/4}\beta^{3/2}}e^{-p^{2}/(2\beta^{2})}\ ,
\end{equation}
and then used  the wave function obtained in this way
to get pseudo-scalar and vector decay constants from
(\ref{fi}), (\ref{fi1}) and (\ref{fi2}).
However, a single harmonic oscillator (HO)
 basis state is not a suitable approximation
for the meson wave function. Namely, lattice simulations
\cite{duncan}
show that heavy-light wave functions fall exponentially with large
$r$ ($\sim e^{-\beta r}$), and therefore HO
wave functions ($\sim e^{-\beta^{2}r^{2}/2}$) cannot be
expected to reflect the correct  dynamics of heavy-light mesons.
If single basis states are used, a much better choice would  be
pseudo-Coulombic (PC) basis states \cite{weniger}
which fall exponentially with large $r$ and
appear to be in a good agreement with the lattice data, as can be
seen  in Figure \ref{wfplt}.

Models such as the one we are using here
are usually solved by diagonalizing the Hamiltonian matrix in a
particular (truncated) basis, with basis states depending
on some variational  parameter \cite{olsson}.
As one increases the number of basis states, the dependence of
eigenvalues and eigenfunctions on the variational parameter should
vanish for the lowest states.
In the case of
QCD-motivated  potentials the solutions obtained with the
PC wave functions converge much more rapidly
with an increase in the number of basis states,  than those obtained
with the HO wave functions. We illustrate that
in Figure \ref{bplt}, by plotting the dependence of energy of the
lowest
$1S$ state on the variational parameter for $N=1$, 5 and 15
basis states, for both PC and HO wave functions.
One can clearly see that the lowest $1S$ HO wave function is not a
very good trial wave function
in a variational calculation of (\ref{srqm}) (with
QCD-motivated potentials).
Furthermore, even if one believes that the $N=1$ HO result
for a state energy is acceptable (it is roughly $50\ MeV$
higher than the exact solution, as can be
seen in Figure \ref{bplt}), that still does not justify
the use of a single HO basis state as a meson wave function.
This issue is clearly important in calculations where
a correct description of meson dynamics is needed, such
as calculations of meson decay constants.
Results obtained by varying a single HO basis state are thus to be
interpreted as non-relativistic estimates of some effective harmonic
oscillator potential, and not as the results of a QCD-motivated
relativistic quark model.

One can now observe that if one uses enough basis states, the
choice of basis wave functions should not matter, and
 pseudo-scalar and vector decay constants should be obtainable from
the relativistic quark model considered here. The problem
is, however, that the $1S$
wave function is divergent
at the spatial origin \cite{amundson}, i.e.,
\begin{equation}
\psi_{1S} \sim r^{-4\alpha_{s}/(3 \pi)}\ .
\label{div}
\end{equation}
The singularity for $r\rightarrow 0$ is related to the singularity of
the short-range Coulomb  potential.
By increasing the number of (usually finite at $r=0$)  basis states,
one is gradually beginning  to see that singularity \cite{jacobs}.
Furthermore, from (\ref{div}) one can see that the degree of divergence
highly depends on the choice of $\alpha_{s}$.
Because of that, one can expect that pseudo-scalar and
vector decay constants cannot be reliably estimated
within the model we are considering.
In Figures \ref{fpsplt} and \ref{fvplt} we demonstrate
the dependence
of the pseudo-scalar ($D$-meson)
and vector ($D^{*}$-meson)
decay constants on the number
of basis states ($N$), for both PC and HO wave functions.
As one can see, for small $N$ both $f_{P}$ and $f_{V_{1/2}}$ are
significantly increasing
with an  increase in $N$.
By including enough basis states, the dependence on $N$ would
eventually vanish.\footnote{Because of the minus sign
in (\ref{fi1}) the results
for $f_{P}$ are better behaved than those for $f_{V_{1/2}}$.
For example, $f_{P}$ obtained with $N=50$ PC states are usually
larger  than those obtained with $N=25$ by only a few $MeV$.
On the other hand, the same increase in $N$
in general leads to increase in $f_{V_{1/2}}$ by several
hundred $MeV$.}
However,  as implied by (\ref{div}), both $f_{P}$ and $f_{V_{1/2}}$
are quite sensitive to the particular choice of
parameters of the model. In our calculations we have observed  that
results obtained with
fixed $N$ can vary  up to a few hundred $MeV$.
Because of that,
we were not able to obtain
reliable estimates of  $f_{P}$
and $f_{V_{1/2}}$ from the model considered in this paper.\footnote{From
Figures \ref{fpsplt} and \ref{fvplt} it should be clear that in the model
considered here
the ratio  $f_{P}/f_{V_{1/2}}$ also cannot be determined with
reasonable errors.}

One possible solution of the problem discussed above
would be to replace the $1/r$ potential with the one-loop
single gluon exchange potential, i.e.,  $\alpha_{s} \rightarrow
\alpha_{s}(r)$.
The $1S$ solution of (\ref{srqm}) in that case
is still divergent, but the
divergence is only logarithmic \cite{amundson}.
This should lead to much more stable results than the ones
shown in Figures \ref{fpsplt} and \ref{fvplt}. These
results should also be much less dependent
on the specific choice of the model parameters. In fact,
such a calculation for $f_{P}$ (for $D$, $D_{s}$,
$B$, and $B_{s}$ mesons)
was already performed by Capstick and Godfrey
in \cite{capstick} using  the model of  \cite{godf}.
The dependence of their results on the number of basis states
was not shown, but
the authors of \cite{capstick} stated that they believed that the  model
overestimates pseudo-scalar decay constants (e.g.,  for
$D$ meson they found $f_{P}=301\ MeV$ with uncertainty
of 20\%).  Even though
it is important to investigate  what really
happens with both $f_{P}$ and $f_{V_{1/2}}$ in
such a model, we shall not consider it in the present paper.

We next  discuss the heavy-light
$P$ and $D$-wave decay constants. While we were not
able to obtain reliable results from (\ref{srqm}) and (\ref{pot})
for the $S$-waves, the situation for $P$ and $D$-waves
is completely different. In Figures \ref{fsplt}, \ref{fpvplt},
\ref{fpv2plt} and \ref{fv2plt} we show the dependence on the
number of basis states ($N$),
for scalar ($S$), two pseudo-vector
($A_{1/2}$ and  $A_{3/2}$),
and vector ($V_{3/2}$) decay constants, respectively.
All the results shown are for the $D^{**}$ mesons.
As one can see in those figures, in general only a few basis
states are needed for results to become independent of $N$, even
though the derivatives of the actual $1P$ and $1D$
wave functions are singular at spatial origin
\cite{jacobs}.\footnote{By
fixing all input parameters, the sensitivity of the decay
constants on the number of basis states was investigated.
To achieve an accuracy
of $0.1$ MeV for $f_{S}$ and $f_{V_{3/2}}$ as little as
10 PC basis states usually were
needed, while to achieve the same accuracy for
$f_{A_{1/2}}$ and $f_{A_{3/2}}$ requires in
general about 50 to 75 PC basis states.}
Furthermore, as $N$ increases the HO results approach the PC
results (always from below) which shows that the difference
between the two basis sets is slowly vanishing.
However, even with 15 basis states (when the state energy obtained  from the
model is essentially equal for both PC and HO wave functions), we can  still
see the difference for $f_{A_{1/2}}$ (Figure \ref{fpvplt})
and for $f_{A_{3/2}}$ (Figure \ref{fpv2plt}). This
reflects the difference in the wave functions obtained from the
two basis sets. The reason why both PC and HO basis states
yield almost the same results for $f_{S}$ (Figure \ref{fsplt}),
even though $0^{+}_{1/2}$ state is also a
$P$-wave, is the minus sign in (\ref{fi3}).
Of course,  because of the much more rapid convergence,
the $PC$ results are to be preferred over the HO results.

Our calculations of $P$ and $D$-wave decay constants
showed that their dependence
on the particular choice of the model parameters
is significantly smaller than the corresponding
dependence of $f_{S}$ and $f_{V_{1/2}}$.
We present the results for $D^{**}$, $D_{s}^{**}$, $B^{**}$, and
$B_{s}^{**}$ mesons in Tables \ref{td}, \ref{tds}, \ref{tb}, and
\ref{tbs}, respectively.\footnote{All results
given in Tables \ref{td} through \ref{tbs} were
obtained with 25 PC basis states, which  was more than
enough for  the accuracy of less than $1\ MeV$ in all
cases considered.} To obtain these results
the effective string tension $b$
of the model was determined from the observed slope of the
$\rho$ trajectory. For a fixed $m_{u,d}$ other parameters
were obtained from the spectrum of
the known heavy-light states.
Experimental meson masses were used in (\ref{fi}) only when their quantum
numbers were unambiguously determined.  Else, we used model predictions
for the appropriate spin-averaged masses, which are also shown
in Tables \ref{td} through \ref{tbs}.

In order to estimate uncertainties introduced by a particular
choice of the constituent mass of $u$ and $d$ quarks,
we have varied $m_{u,d}$ in the range from $150\ MeV$ to $350\ MeV$.
For a given $m_{u,d}$,  by adjusting $c$
we have also varied constituent heavy quark
masses in the range of about $400\ MeV$
(e.g.,  $m_{c}$ was varied in the range from about $1.3\ GeV$
to about $1.7\ GeV$). We emphasize that a good description
of the spin-averaged heavy-light meson spectrum was always
maintained.

Results for the decay constants
obtained in this way depend on the assumption
for the unknown spin-averaged mass  of
$D_{0}$  and $D'_{1}$ mesons
($0^{+}_{1/2}$ and $1^{+}_{1/2}$ states). To take  into account
ambiguities introduced in our results in that way,
we have repeated all calculations for this unknown mass
in the range from $2200\ MeV$ to $2450\ MeV$.
Errors quoted in Tables \ref{td} through \ref{tbs}
reflect the uncertainty due to the unknown $P$-wave mass,
as well as the  uncertainties related
to the choice of constituent quark masses discussed above.

As one can see from those tables, in spite of the fact
that our calculations are performed for a broad range of
constituent quark masses, and also for a wide range of
the unknown $P$-wave mass,
as long as a good description
of  the observed heavy-light meson spectrum is maintained,
the $P$ and $D$-wave heavy-light decay constants
are all predicted rather precisely. It is also interesting
to observe that the decay constants of strange
$0_{1/2}^{+}$,
$1_{1/2}^{+}$, and
$1_{3/2}^{-}$ states are slightly smaller
than those of the corresponding non-strange states.
The main reason is (besides the  $1/\sqrt{M}$ dependence
of (\ref{fi})) the light quark dependence of
(\ref{fi3}), (\ref{fi4}) and (\ref{fi6}).
On the other hand, (\ref{fi5}) does not depend on
the light quark mass, so that $\sqrt{(m_{q}+E_{q})/E_{q}}$
factor in (\ref{fi}) plays a much more significant role,
and as a result $f_{A_{3/2}}$ for the strange
states are larger than the ones for non-strange states.
Also note that $f_{S}$ for
$B_{0}$ and $B_{s0}$ are larger than those of the corresponding
$D_{0}$ and $D_{s0}$ states, while it is the other way
around in the case of $f_{A_{1/2}}$.
The reason  for this are  the  minus
and plus signs in (\ref{fi3}) and (\ref{fi4}), respectively.
Finally, the fact that
$1_{3/2}^{+}$ and
$1_{3/2}^{-}$ $B^{**}$ states have decay constants
smaller than those of the corresponding $D^{**}$ states,
can be easily explained with the
$1/(m_{\overline{Q}}+E_{\overline{Q}})$
dependence of (\ref{fi5}) and (\ref{fi6}).

\section{Conclusion}
\label{conc}

In this paper we have examined decay constants
of  heavy-light mesons within the mock-meson approach
\cite{yaou,ruiz,isgu,hayne}.
We obtained all the relevant expressions
in the $j$-$j$ coupling scheme. For numerical estimates
we employed  a simple  and widely used relativistic quark model
\cite{godf,durand,lichtenberg,jacobs}. It
is based on a spinless Salpeter equation
with QCD-motivated Coulomb-plus-linear potential.
The effective string tension is   chosen so that the Regge structure
of the model in the light-light limit  is consistent
with  experiment,  and other parameters are based on the
good description of the known spin-averaged
heavy-light  meson masses.

Due to the singular nature of the $L=0$  wave functions at spatial
origin \cite{amundson}, we were not able to obtain
reliable estimates of pseudo-scalar and vector decay constants.
On the other hand, even though we have
allowed for large variations of input parameters,
our results show that the model predicts a
rather narrow range  for all lowest $P$ and $D$-wave heavy-light
decay constants.

Such precisely predicted decay constants allow us to estimate
the $D_{(s)}^{**}$ production fractions in $b$
decays governed by the $b\rightarrow c\overline{c}s'$ transitions
under the factorization assumption. Quantitative predictions
regarding the interference
of color-allowed and color-suppressed amplitudes in
$B^{-}\rightarrow D^{**0}\{\pi^{-},\rho^{-},a_{1}^{-},\ldots\}$ modes
can now be formulated. These and some other consequences of our   findings
will be discussed elsewhere \cite{dunietz}.

\acknowledgments

We thank J. F. Amundson, E. Eichten  and D. Zeppenfeld for   discussions.
S. Veseli would also like to thank the theory group for hospitality
during his visit to Fermilab.
This work was supported in part by the U.S. Department of Energy
under Contracts No. DE-AC02-76CH03000 and
 DE-FG02-95ER40896, and in part by the University
of Wisconsin Research Committee with funds granted by the Wisconsin
Alumni
Research Foundation.

\begin{table}
\caption{Relativistic quark
model  predictions compared to  experimental spin-averaged
heavy-light meson masses. Parameters of the model
are
$m_{u,d}=0.300\ GeV$, $m_{s}=0.483\ GeV$,
$m_{c}= 1.671\ GeV$, $m_{b}=5.121\ GeV$,
$\alpha_{s}= 0.498$, $b=0.142\ GeV^{2}$, and $c=-0.350\ GeV$.
 The unknown
$D_{0}$  and $D'_{1}$ mesons
($0^{+}_{1/2}$ and $1^{+}_{1/2}$ states) were assumed to
have a spin-averaged mass of 2400$\ MeV$. Heavy quark symmetry
arguments then lead to the spin-averaged mass of 2502 $MeV$
for the corresponding $D_{s0}$ and $D'_{s1}$ mesons.}
\label{spectrum}
\begin{tabular}{|lcccc|}
Meson & State & Experiment & Theory & Error \\
      &       &  $[MeV]$    & $[MeV]$ & $[MeV]$ \\
\hline
$\begin{array}{l}
D(1867) \\
D^{*}(2009)
\end{array} $ &$  \left. \begin{array}{c}
		0^{-}_{1/2} \\
		1^{-}_{1/2}
		\end{array} \right\} $ &  $1S(1974) $ & 1971 & $-3$
\\
$\begin{array}{l}
D_{0}(\sim 2400) \\
D'_{1}(\sim 2400) \\
D_{1} (2425) \\
D_{2}^{*}(2459)
\end{array} $ &$  \left. \begin{array}{c}
		0^{+}_{1/2} \\
		1^{+}_{1/2} \\
		1^{+}_{3/2} \\
		2^{+}_{3/2}
		\end{array} \right\} $ &  $1P(2431) $ & 2434 & $+3$
\\
$\begin{array}{l}
D_{s}(1969) \\
D_{s}^{*}(2112)
\end{array} $ &$  \left. \begin{array}{c}
		0^{-}_{1/2} \\
		1^{-}_{1/2}
		\end{array} \right\} $ &  $1S(2076) $ & 2079 & $+3$
\\
$\begin{array}{l}
D_{s0}(\sim 2502) \\
D'_{s1}(\sim 2502) \\
D_{s1} (2535) \\
D_{s2}^{*}(2573)
\end{array} $ &$  \left. \begin{array}{c}
		0^{+}_{1/2} \\
		1^{+}_{1/2} \\
		1^{+}_{3/2} \\
		2^{+}_{3/2}
		\end{array} \right\} $ &  $1P(2540) $ & 2537 & $-3$
\\
$\begin{array}{l}
B(5279) \\
B^{*}(5325)
\end{array} $ &$  \left. \begin{array}{c}
		0^{-}_{1/2} \\
		1^{-}_{1/2}
		\end{array} \right\} $ &  $1S(5314) $ & 5314 &  $+0$
\\
$\begin{array}{l}
B_{s}(5374) \\
B_{s}^{*}(5421)
\end{array} $ &$  \left. \begin{array}{c}
		0^{-}_{1/2} \\
		1^{-}_{1/2}
		\end{array} \right\} $ &  $1S(5409) $ & 5409 &  $-0$
\end{tabular}
\end{table}

\begin{table}
\caption{Decay constants of heavy-light $D^{**}$ states, as obtained
from
the relativistic quark model. Whenever possible we used
experimental meson masses. If these were unknown,
we used model predictions
for the spin-averaged masses.}
\label{td}
\begin{tabular}{|lcc|}
Meson & State & $f_{i}$ \\
      &       & $[MeV]$ \\
\hline
$D(1867)$  & $1S,\ 0^{-}_{1/2}$ & not reliable  \\
$D^{*}(2009)$  & $1S,\ 1^{-}_{1/2}$ & not reliable \\
$D_{0}(2410\pm 40)$  & $1P,\ 0^{+}_{1/2}$ & $139\pm 30$  \\
$D'_{1}(2410\pm 40)$  & $1P,\ 1^{+}_{1/2}$ & $251\pm  37$  \\
$D_{1}(2425)$  & $1P,\ 1^{+}_{3/2}$ & $77\pm 18$  \\
$D''_{1}(2700\pm 55)$  & $1D,\ 1^{-}_{3/2}$ & $48 \pm 7$
\end{tabular}
\end{table}

\begin{table}
\caption{Decay constants of heavy-light $D_{s}^{**}$ states, as
obtained from
the relativistic quark model. Whenever possible we used
experimental meson masses. If these were unknown,
we used model predictions
for the spin-averaged masses.}
\label{tds}
\begin{tabular}{|lcc|}
Meson & State & $f_{i}$ \\
      &       & $[MeV]$ \\
\hline
$D_{s}(1969)$  & $1S,\ 0^{-}_{1/2}$ & not reliable  \\
$D_{s}^{*}(2112)$  & $1S,\ 1^{-}_{1/2}$ & not reliable  \\
$D_{s0}(2510\pm 45)$  & $1P,\ 0^{+}_{1/2}$ & $110\pm 18$  \\
$D'_{s1}(2510\pm 45)$  & $1P,\ 1^{+}_{1/2}$ & $233\pm 31$  \\
$D_{s1}(2535)$  & $1P,\ 1^{+}_{3/2}$ & $87\pm 19$  \\
$D''_{s1}(2795\pm 55)$  & $1D,\ 1^{-}_{3/2}$ & $45\pm 6$
\end{tabular}
\end{table}

\begin{table}
\caption{Decay constants of heavy-light $B^{**}$ states, as obtained  from
the relativistic quark model. Whenever possible we used
experimental meson masses. If these were unknown,
we used model predictions
for the spin-averaged masses.}
\label{tb}
\begin{tabular}{|lcc|}
Meson & State & $f_{i}$  \\
      &       & $[MeV]$  \\
\hline
$B(5279)$  & $1S,\ 0^{-}_{1/2}$ & not reliable  \\
$B^{*}(5325)$  & $1S,\ 1^{-}_{1/2}$ & not reliable  \\
$B_{0}(5765\pm 60)$  & $1P,\ 0^{+}_{1/2}$ & $162\pm 24$  \\
$B'_{1}(5765\pm 60)$  & $1P,\ 1^{+}_{1/2}$ & $206\pm 29$  \\
$B_{1}(5765\pm 60)$  & $1P,\ 1^{+}_{3/2}$ & $32\pm 10$  \\
$B''_{1}(6040\pm 70)$  & $1D,\ 1^{-}_{3/2}$ & $18\pm 3$
\end{tabular}
\end{table}

\begin{table}
\caption{Decay constants of heavy-light $B_{s}^{**}$ states, as
obtained from
the relativistic quark model. Whenever possible we used
experimental meson masses. If these were unknown,
we used model predictions
for the spin-averaged masses.}
\label{tbs}
\begin{tabular}{|lcc|}
Meson & State & $f_{i}$ \\
      &       & $[MeV]$  \\
\hline
$B_{s}(5374)$  & $1S,\ 0^{-}_{1/2}$ & not reliable  \\
$B_{s}^{*}(5421)$  & $1S,\ 1^{-}_{1/2}$ & not reliable  \\
$B_{s0}(5860\pm 65)$  & $1P,\ 0^{+}_{1/2}$ & $146\pm 19$  \\
$B'_{s1}(5860\pm 65)$  & $1P,\ 1^{+}_{1/2}$ & $196\pm 26$  \\
$B_{s1}(5860\pm 65)$  & $1P,\ 1^{+}_{3/2}$ & $36\pm 10$  \\
$B''_{s1}(6130\pm 75)$  & $1D,\ 1^{-}_{3/2}$ & $17\pm 3$
\end{tabular}
\end{table}

\begin{figure}
\caption{Comparison of the pseudo-Coulombic (PC,
$R_{1S}(r) = 2\beta^{3/2}
e^{-\beta r}$), and the harmonic
oscillator (HO, $R_{1S}(r) = 2\beta^{3/2}/\pi^{1/4}
e^{-\beta^{2}r^{2}/2}$), $1S$ configuration space wave functions
 with the lattice data
\protect\cite{duncan}. For both PC and HO wave functions we used
$\beta = 0.40 \ GeV$.}
\label{wfplt}
\end{figure}

\begin{figure}
\caption{Convergence of the $1S$ state mass of (\protect\ref{srqm})
and (\protect\ref{pot}),
with $m_{1}=m_{u,d}=0.300\ GeV$, $m_{2}= m_{c}=1.671\ GeV$,
$b = 0.142\ GeV^{2}$, $ c = -0.350\ GeV$, and $\alpha_{s} = 0.498$.
Pseudo-Coulombic (PC, full lines) and harmonic
oscillator (HO, dashed lines) wave functions  with
$N=1$, 5, and 15 basis states.}
\label{bplt}
\end{figure}

\begin{figure}
\caption{Dependence of the pseudo-scalar
($D$ meson, $0^{-}_{1/2}$ state) decay constant $f_{P}$ on the number
$(N)$ of pseudo-Coulombic (PC), and harmonic oscillator (HO) basis
states.
We have used parameters given in the text,
i.e.,  $m_{1}=m_{u,d}=0.300\ GeV$, $m_{2}= m_{c}=1.671\ GeV$,
$b = 0.142\ GeV^{2}$, $ c = -0.350\ GeV$, and $\alpha_{s} = 0.498$.}
\label{fpsplt}
\end{figure}

\begin{figure}
\caption{Dependence of the vector
($D^{*}$ meson, $1^{-}_{1/2}$ state) decay constant $f_{V_{1/2}}$
on the number
$(N)$ of pseudo-Coulombic (PC), and harmonic oscillator (HO) basis
states.
We have used
$m_{1}=m_{u,d}=0.300\ GeV$, $m_{2}= m_{c}=1.671\ GeV$,
$b = 0.142\ GeV^{2}$, $ c = -0.350\ GeV$, and $\alpha_{s} = 0.498$.}
\label{fvplt}
\end{figure}

\begin{figure}
\caption{Dependence of the scalar
($D_{0}$ meson, $0^{+}_{1/2}$ state) decay constant $f_{S}$
on the number
$(N)$ of pseudo-Coulombic (PC), and harmonic oscillator (HO) basis
states.
We have used
$m_{1}=m_{u,d}=0.300\ GeV$, $m_{2}= m_{c}=1.671\ GeV$,
$b = 0.142\ GeV^{2}$, $ c = -0.350\ GeV$, and $\alpha_{s} = 0.498$.}
\label{fsplt}
\end{figure}

\begin{figure}
\caption{Dependence of the pseudo-vector
($D'_{1}$ meson, $1^{+}_{1/2}$ state) decay constant $f_{A_{1/2}}$
on the number
$(N)$ of pseudo-Coulombic (PC), and harmonic oscillator (HO) wave
functions.
We have used
$m_{1}=m_{u,d}=0.300\ GeV$, $m_{2}= m_{c}=1.671\ GeV$,
$b = 0.142\ GeV^{2}$, $ c = -0.350\ GeV$, and $\alpha_{s} = 0.498$.}
\label{fpvplt}
\end{figure}

\begin{figure}
\caption{Dependence of the pseudo-vector
($D_{1}$ meson, $1^{+}_{3/2}$ state) decay constant $f_{A_{3/2}}$
on the number
$(N)$ of pseudo-Coulombic (PC), and harmonic oscillator (HO) wave
functions.
We have used
$m_{1}=m_{u,d}=0.300\ GeV$, $m_{2}= m_{c}=1.671\ GeV$,
$b = 0.142\ GeV^{2}$, $ c = -0.350\ GeV$, and $\alpha_{s} = 0.498$.}
\label{fpv2plt}
\end{figure}

\begin{figure}
\caption{Dependence of the vector
($D''_{1}$ meson, $1^{-}_{3/2}$ state) decay constant $f_{V_{3/2}}$
on the number
$(N)$ of pseudo-Coulombic (PC), and harmonic oscillator (HO) wave
functions.
We have used
$m_{1}=m_{u,d}=0.300\ GeV$, $m_{2}= m_{c}=1.671\ GeV$,
$b = 0.142\ GeV^{2}$, $ c = -0.350\ GeV$, and $\alpha_{s} = 0.498$.}
\label{fv2plt}
\end{figure}

\newpage

\begin{figure}[p]
\epsfxsize = 5.4in
\centerline{\vbox{\epsfbox{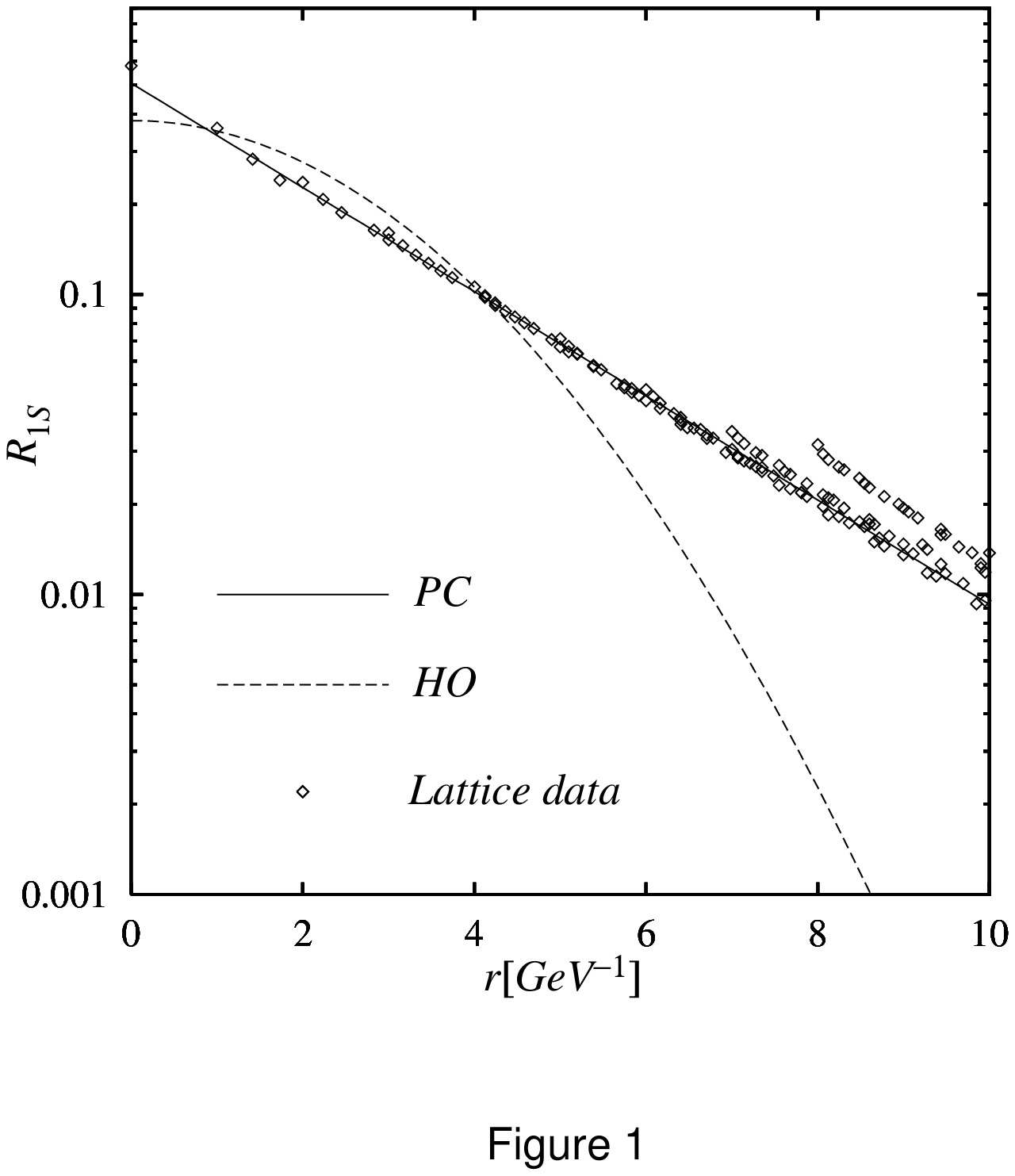}}}
\end{figure}

\begin{figure}[p]
\epsfxsize = 5.4in
\centerline{\vbox{\epsfbox{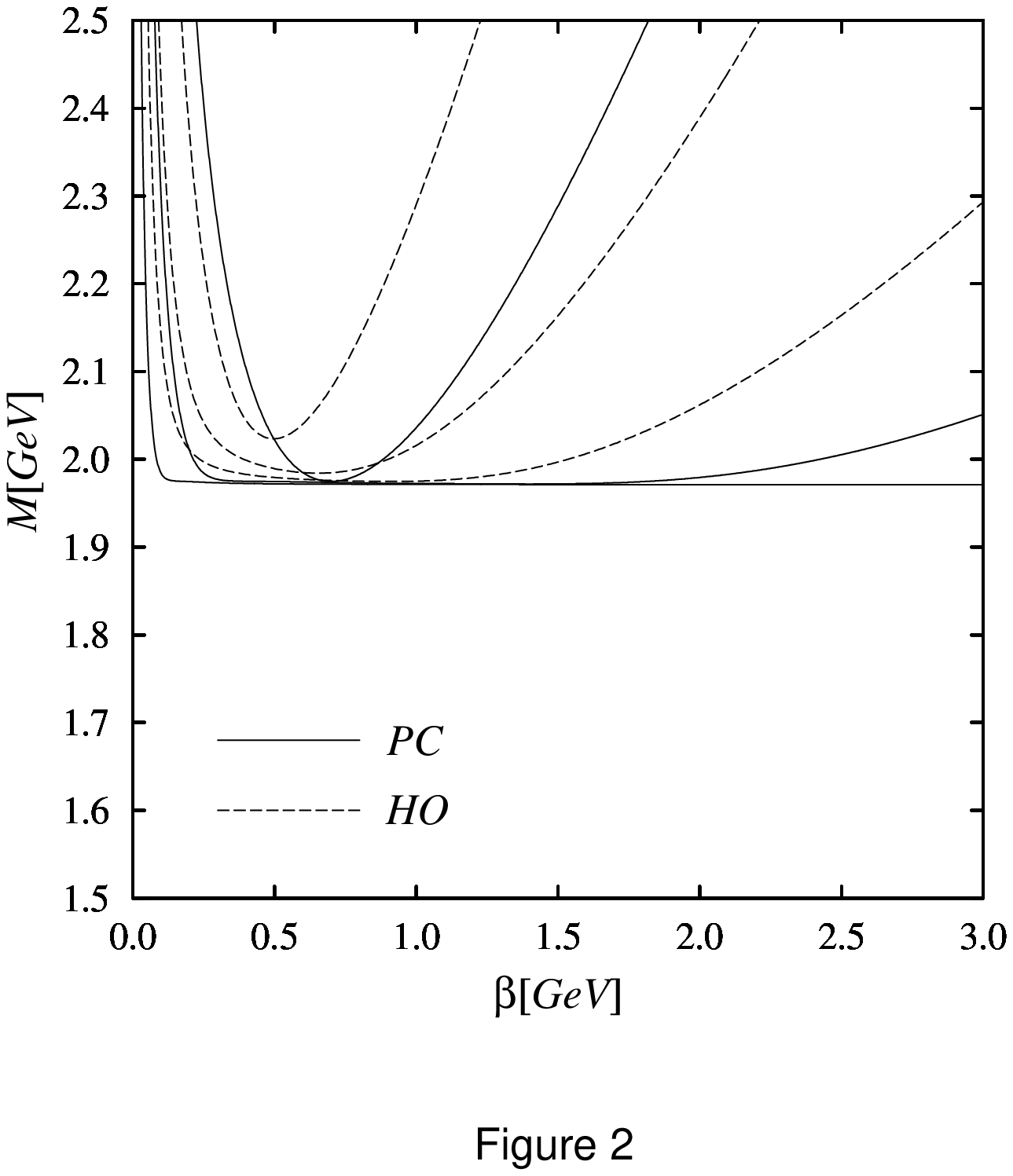}}}
\end{figure}

\begin{figure}[p]
\epsfxsize = 5.4in
\centerline{\vbox{\epsfbox{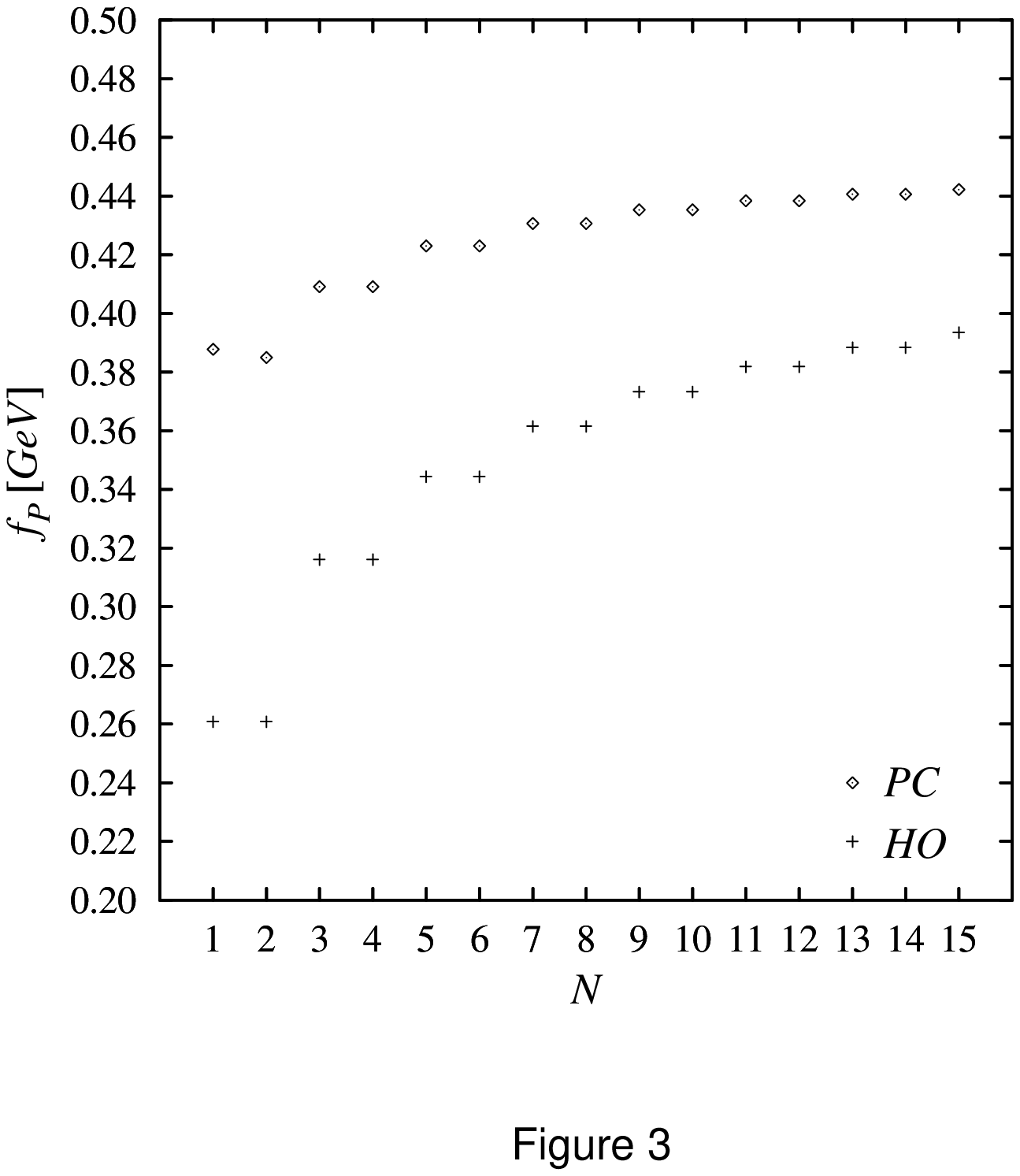}}}
\end{figure}

\begin{figure}[p]
\epsfxsize = 5.4in
\centerline{\vbox{\epsfbox{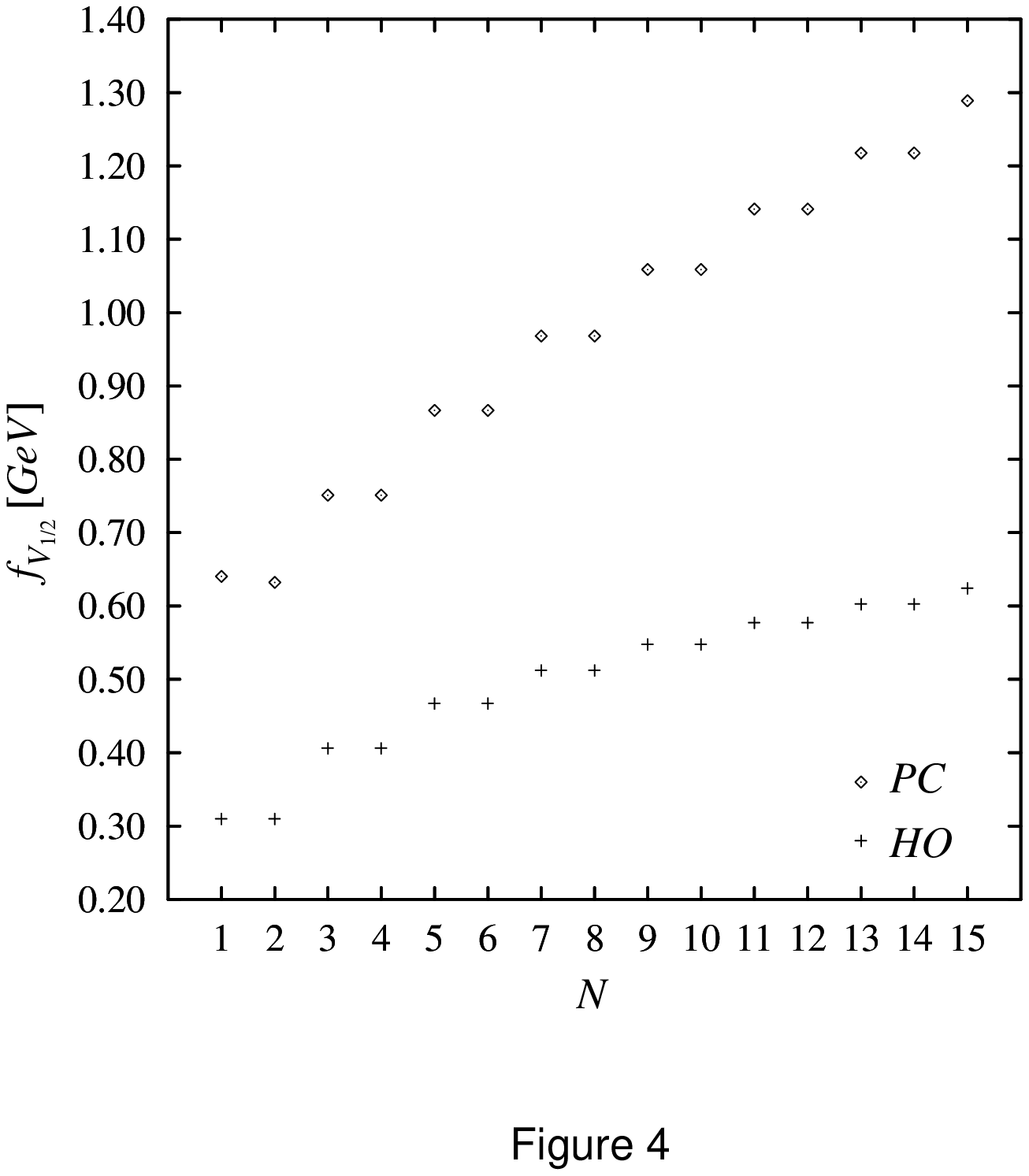}}}
\end{figure}

\begin{figure}[p]
\epsfxsize = 5.4in
\centerline{\vbox{\epsfbox{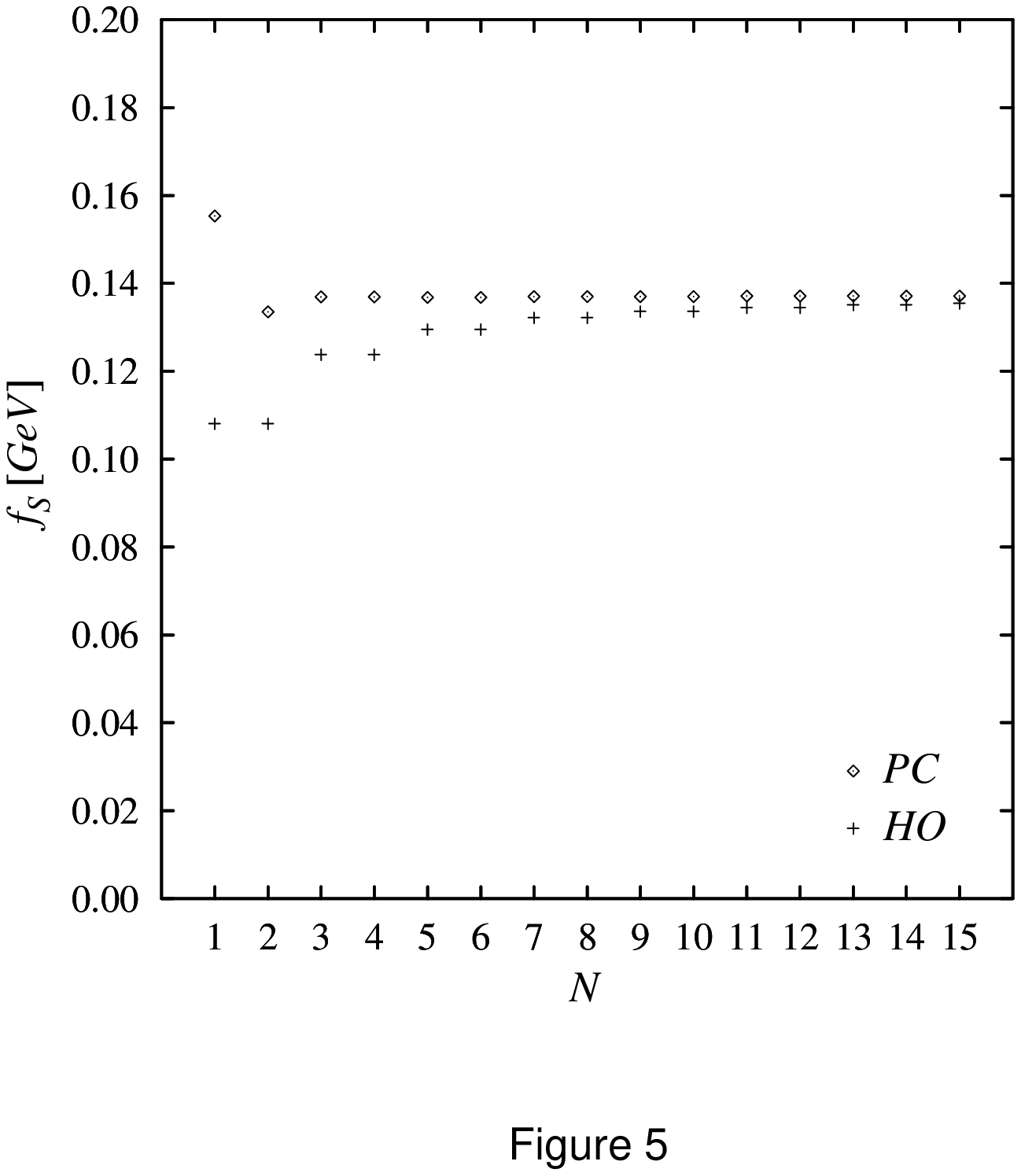}}}
\end{figure}

\begin{figure}[p]
\epsfxsize = 5.4in
\centerline{\vbox{\epsfbox{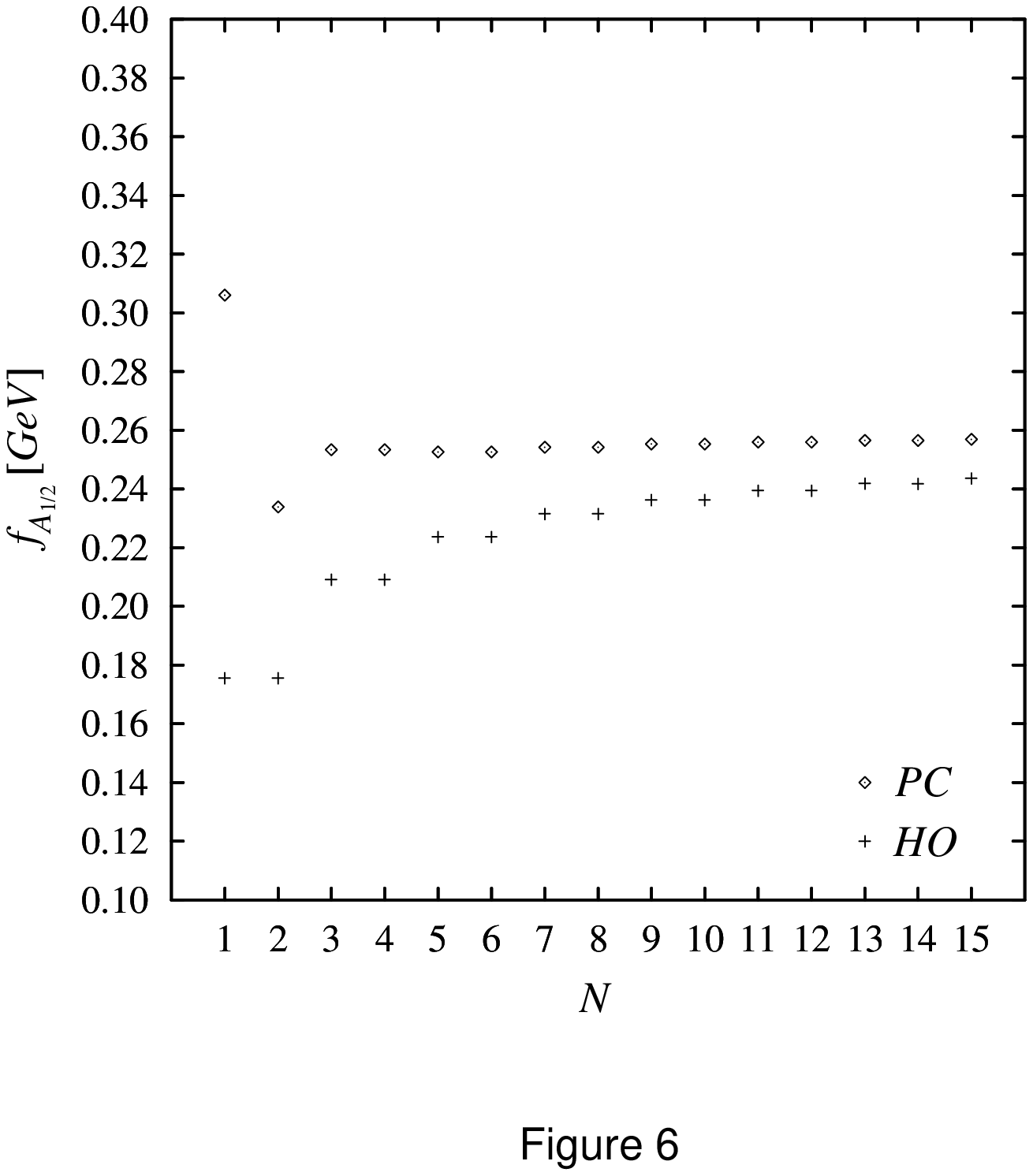}}}
\end{figure}

\begin{figure}[p]
\epsfxsize = 5.4in
\centerline{\vbox{\epsfbox{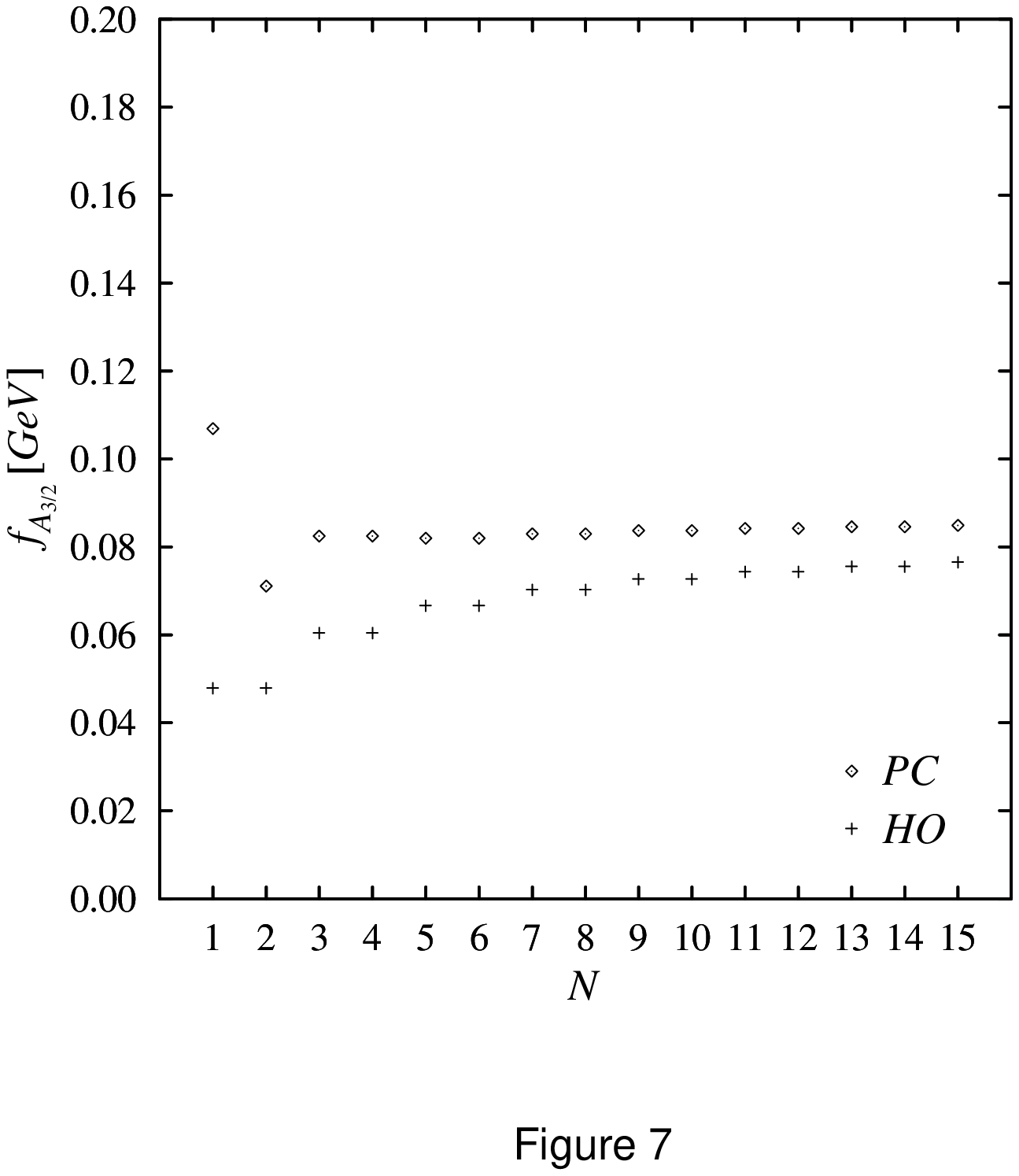}}}
\end{figure}

\begin{figure}[p]
\epsfxsize = 5.4in
\centerline{\vbox{\epsfbox{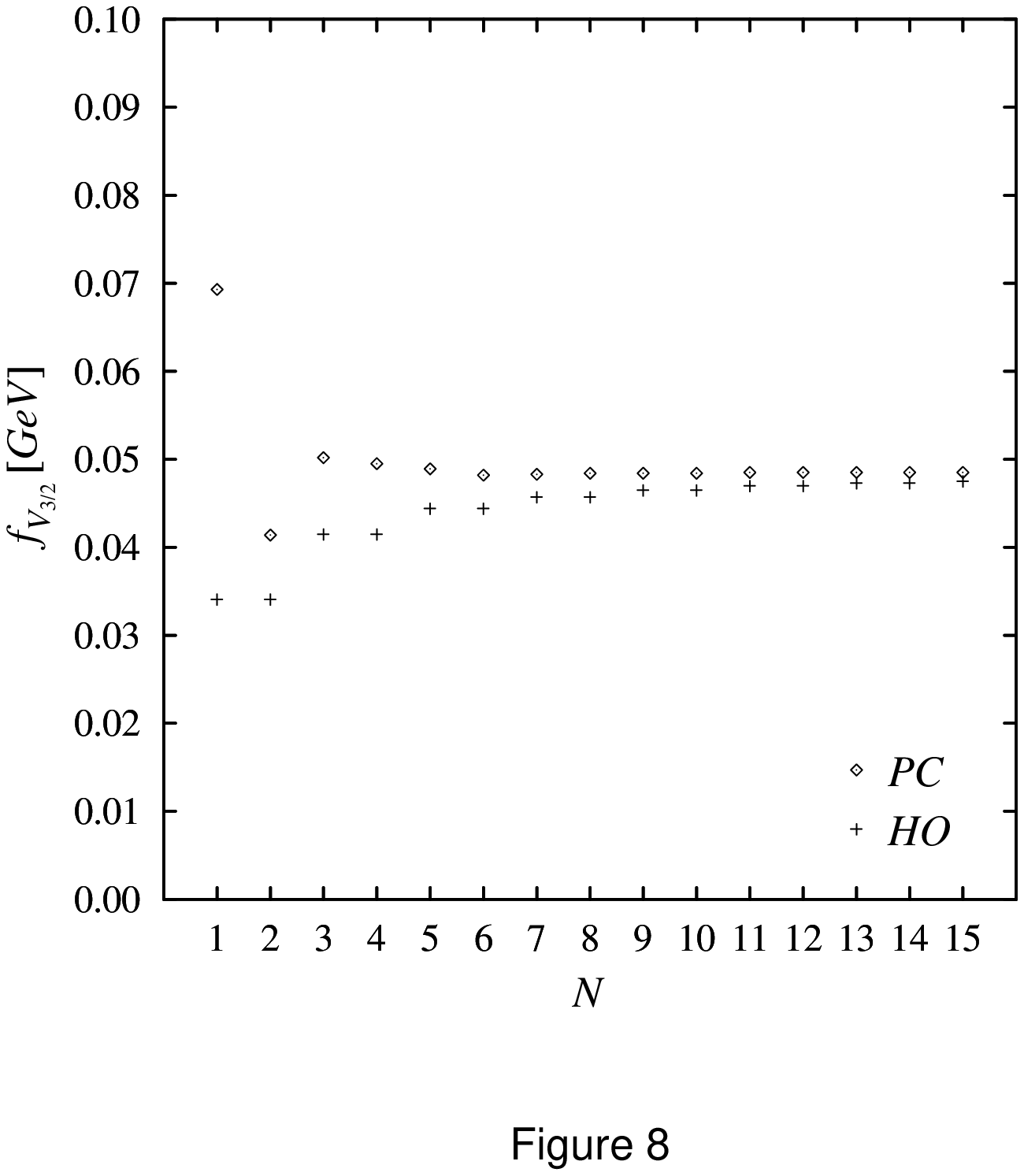}}}
\end{figure}

\end{document}